# THE SPATIAL NEAREST NEIGHBOR SKYLINE QUERIES

Nasrin Mazaheri Soudani[1] and Ahmad Baraani Dastgerdi[2]

[1]Department of Computer Engineering, Isfahan University, Isfahan, Iran
`nasrinmazaheri@eng.ui.ir`
[2]Department of Computer Engineering, Isfahan University, Isfahan, Iran
`ahmadb@eng.ui.ir`

*ABSTRACT*

*User preference queries are very important in spatial databases. With the help of these queries, one can found best location among points saved in database. In many situation users evaluate quality of a location with its distance from its nearest neighbor among a special set of points. There has been less attention about evaluating a location with its distance to nearest neighbors in spatial user preference queries. This problem has application in many domains such as service recommendation systems and investment planning. Related works in this field are based on top-k queries. The problem with top-k queries is that user must set weights for attributes and a function for aggregating them. This is hard for him in most cases. In this paper a new type of user preference queries called spatial nearest neighbor skyline queries will be introduced in which user has some sets of points as query parameters. For each point in database attributes are its distances to the nearest neighbors from each set of query points. By separating this query as a subset of dynamic skyline queries $N^2S^2$ algorithm is provided for computing it. This algorithm has good performance compared with the general branch and bound algorithm for skyline queries.*

## 1. INTRODUCTION

In many situations for decision making, users need select one or more data from database in accordance with their interest. The selected data must meet their desired constraints. For example suppose in a database about a shoreline city, information of its hotels such as cost and distance of each hotel from beach has been saved. A user wants to select a hotel with less cost and distance to beach. User hasn't accurate asked (for example cost of hotel below 100$ and distance to beach less than 1Km is accurate asked) but wants to find a set of data that are closer to their own interests. Such constraints called soft constraints and queries about these problems called user preference queries [1].

There are two basic queries for these problems. In the first type of queries that are known to top-k, each of data attribute based on their importance to user gives a weight. The score of the data is computed by multiplying its values with the corresponding weights and aggregating them by a function. This query retrieves the k data with the highest scores [2]. In the second type of queries that are known to skyline, the set of all data that no other data dominate them are retrieved. A data dominate another if and only if for all attributes is better than or equals and for at least one

International Journal of Database Management Systems ( IJDMS ) Vol.3, No.4, November 2011

attribute is better than the other. Indeed, in this way all data that aren't worse than any other data in database are retrieved [3].

User preference queries are very important in spatial databases. Spatial data in addition to non-spatial data can be stored in these databases. With the help of these queries, user can find best places in database according to their interest.

For many application users evaluate quality of a location with its distance from its nearest neighbor among special set of points. For example suppose a user wants to find a hotel for rest that is near to a restaurant and a coffee shop. So he considers coffee shops and restaurants as two query point sets and evaluates hotels with their distances to nearest coffee shop and restaurant. Less attention has been about subject distance of a point to its nearest neighbors as preference of the user.

Related works in this field are in based on top-k queries. In top-k queries setting a weight for each attribute and a scoring function for aggregating attributes are hard for user. Indeed user is more willing to ask for the skyline first in order to get the "big picture" and then apply a top-k query to get more specific results. So the use of skyline query with considering the distances of points to their nearest neighbors is subject that less attention has been on it [3].

Distance of the point to its nearest neighbor is a dynamic attribute for that point. Dynamic attributes are not directly stored in database but according to data from database can be calculated. Existence algorithms for respond skyline queries such as branch and bound algorithm don't have good performance for dynamic skyline queries and in based on type of dynamic attributes, many optimizations can be performed on them[4].

In this paper a new type of dynamic skyline query called spatial nearest neighbor skyline query is introduced. In these queries, user has some set of query points. For each point in database, attributes are its distances to nearest neighbors from each set of query points. All points that there is a point better than them according to all attributes are deleted and the rest are returned as the answer.

This query is the subset of dynamic skyline queries. By separating it we are trying to provide a more efficient algorithm for computing it. This algorithm that called $N^2S^2$ is in based on branch and bound algorithm and such as branch and bound algorithm supposes there is R tree index on data but this solution can be extended to all kind of data partitioning methods on spatial databases.

In this algorithm by considering the density of the query points in high levels of the index instead of accurate distance to nearest neighbors, a better view with less cost is obtained about nodes of tree may be more promising. Also with storing nodes that are used to find nearest neighbors of parent nodes and using of them for finding nearest neighbor of lower level nodes of tree, many extra fetch of nodes will be avoided.

The rest of the paper is organized as follows: Section 2 at first reviews the related works in this field and then briefly expresses general skyline queries and the branch and bound algorithm for answering them. Section 3 presents formal definition of spatial nearest neighbor skyline queries. Section 4 expresses $N^2S^2$ algorithm for answering this kind of queries. Section 5 evaluates this algorithm comparing general branch and bound algorithm for skyline queries. Finally, Section 6 concludes the paper with a summary of our results and future works.

66

## 2. RELATED WORKS

Previous works in field of user preference queries in spatial databases can be divided into two categories. Papers that are based on top-k queries and ones are based on skyline queries. However more related works with issues presented in this study are in the first category. Spatial skyline queries have been developed only for one type of user preference problems. In some papers such as [5], a combination of both queries is used for spatial user preference problems.

More associated work with this paper is [6]. A type of spatial user preference queries is presented in it that ranks objects based on the qualities of features in their spatial neighborhood and retrieves the K objects in set with the highest ranks. In fact in [6] user has some sets of query points. Each query point in based on its quality has a score between 0 and 1. For each object in database scores of nearest neighbors from each set of query points are aggregated with a function such as sum and object is ranked in based on result of this function. K points that have highest rank are retrieved as query results.

In [6], the distances of objects from their nearest neighbors aren't considered as important factor for retrieving results. So for example an object that its nearest neighbor is far from it may have high rank. This method use top-k queries so user should present an appropriate ranking function.

Writers of [6] extended their works in [7]. In which, in addition to the nearest neighbor, the scores of the higher level neighbors are considered for ranking objects with the condition that the score of i'th nearest neighbor is divided into $2^i$. The problem with this method is that it doesn't consider concept of distance exactly and so for example it is not important that how much the distance of an object with its nearest neighbor is or how much the distance of the first and the second nearest neighbors is.

Skyline queries are used for special kind of spatial user preference queries called location dependent skyline[8,9], multi source skyline [10] and spatial skyline [11] queries. In this kind of queries if an object related to all query points is farer than another object, it will be deleted from results and the rest of objects are retrieved as query results.

The main problem of skyline queries is that if the number of attributes of an object is high and the changes of this attributes are in reverse order, the number of results will be too many that no longer offer any interesting insight[12]. So in spatial skyline queries if the number of query points is high and they are scattered, the number of results will be extremely high. In [11] demonstrated that all points in convex hull of query points, are skyline. So if convex hull of query points equals to the total space, all data objects are skyline. This problem is solved with considering distances to nearest neighbors from each set of query points instead of distances to each query points. in presented query, because for each set of query points consider one dynamic attribute for data objects and because the number of query sets is usually low in application examples, the use of skyline query is appropriate and the number of results would not be out of the user control. On the other hand the distance to nearest neighbor of a set of query points is usually independent from the distance to nearest neighbor of the other set and for this independency the number of results is lower than when the dynamic attributes are anti-correlated to each other [12].

Related to existence or not existence of spatial index on data, different algorithms for replying skyline queries are presented. Famous index on spatial data is R-tree's family. Algorithms that presented with use of this index usually are in based on branch and bound method [4,7,11]. In these algorithms the search starts from the root node of the tree. Tree is traversed and none promising nodes are pruned. Algorithms that are not based on index use concepts such as Voronoi diagram[11,13]. Algorithm presented in this paper is in based on branch and bound method.

In the next subsection branch and bound algorithm is briefly expressed to help better understand our presented algorithm.

## 2.1. GENERAL SKYLINE QUERIES AND THE BRANCH AND BOUND ALGORITHM (BBS)

Given a set of data object P that each object has m attributes, general skyline query retrieves objects such as pi in P that doesn't exist any pj in P that dominates it. A data object dominates another data object if for all attributes is better than or equals and for at least one attribute is better than the other. In this query all objects that may be best for user are retrieved. Indeed this query retrieves results of all top-1 queries. It helps users to get a big picture of the interesting options[3].

One of the best algorithms for responding skyline queries is BBS algorithm presented in [4]. In this algorithm each data object is supposed as a point in m dimensional space. There is a data partitioning method such as R-Tree on data. Search starts from the root node of the tree. A priority queue is used to store tree nodes. At first the root node of the tree is inserted into the priority queue. From then until the queue becomes empty at each stage, a more promising node is removed from the priority queue. If this node is a leaf it will be a skyline point but if this is an intermediate node and not dominated by any found skyline points, it will be expanded and all children that are not dominated by any found skyline points are inserted into priority queue.

Promising criterion for a point is the sum of its attributes or coordinates. Whatever the sum is less, data is more promising. For intermediate nodes, promising criterion is the sum of coordinates of the point with minimum coordinates in minimum bounding rectangle of node.

This algorithm can be used for dynamic skyline queries. In these queries there are m function that take coordinates of a data object as parameters and map it to new m dimensional coordinates. Then the skyline query is applied on the new coordinates.

When applying algorithm on dynamic skyline query, because the dynamic attributes are not saved in database, they should be calculated for any nodes that considered for the first time. According to type of dynamic attributes, many optimizations can do on the BBS algorithm.

## 3. FORMAL PROBLEM DEFINITION

Spatial nearest neighbor skyline query is a subset of skyline query. In skyline query attributes of an object doesn't have any limitation and can be static or dynamic but in spatial nearest neighbor skyline query, attributes of an object are its distances to its nearest neighbors from each sets of query points.

### 3.1. THE SPATIAL NEAREST NEIGHBOR SKYLINE QUERIES

If a set P that contains user interesting data points (for example the set of hotels) and m sets $Q_1, Q_2, \ldots, Q_m$ (for example two sets of coffee shops and restaurants) of user desired query points are given, Spatial nearest neighbor skyline query retrieves all points such as $p_i$ in P that aren't dominated by any other point $p_j$ in P. point $p_j$ dominate $p_i$ if and only if all attributes of $p_j$ is smaller than or equal and for at least one attribute is smaller than to the corresponding attributes of $p_i$. Each point p∈P, has m attributes

$f_j = \min_{1 \leq t \leq n_j}\{D(p, q_{tj})\}$   $1 \leq j \leq m$, $q_{tj} \in Q_t$

In this definition D(.,.) is Euclidean distance between two points. Indeed, dynamic attributes in this definition are distances to nearest neighbors from each set of query points.

## 4. $N^2S^2$ ALGORITHM

### 4.1. PROBLEMS WITH BBS ALGORITHM

If BBS algorithm for dynamic skyline queries is used for spatial nearest neighbor skyline query, when a node is removed from priority queue, its nearest neighbors from each set of query points should be calculated and in based on it decided for that node. Suppose there are separate R tree indexes on each sets of query points. For finding nearest neighbor (from a set of query points) for each child of a node, search should be started from the root of the query points R-tree index, almost the same route that went for finding nearest neighbor of parent node. So fetching a lot of nodes for finding nearest neighbor are repeated. Therefore by storing the nodes of query points R tree that are used to find nearest neighbor of parent node and continue the search from them, we can prevent many of the additional fetches.

On the other hand for higher level and close to the root nodes of data points R tree, don't need the exact distance from its nearest neighbor but only need an overview of the density of query points around the MBR[1] of the node to recognize the priority of it. Therefore the search for finding nearest neighbor need not continue to leaves of query points R tree but nearest MBRs at levels above the leaves can be found and the distances to these MBRs is considered. Therefore many additional fetches of nodes can be avoided.

### 4.2. DISTANCE BETWEEN TWO MBRS

Before presenting algorithm, appropriate metrics for calculating the distance between two MBRs are needed. There are different metrics for the distance of two MBRs. in this paper two metrics that are introduced in [15] will be used. The first is MinMindist that is the minimum distance between any points in two MBRs and the other is MaxMaxdist that is the maximum distance between any points in two MBRs. in figure 1 this two metrics are shown for two MBRs. this two metrics are used for pruning R tree in $N^2S^2$ algorithm.

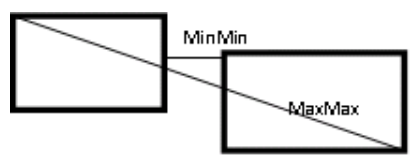

Figure 1. Distance between two MBRs

### 4.3. THE PROCEDURE OF $N^2S^2$ ALGORITHM

$N^2S^2$ Algorithm is in based on BBS Algorithm in which both problems that discussed in BBS algorithm with spatial nearest neighbor skyline queries are solved. The search is started from the root of the data points R tree. For each node of the data points R tree that is fetched, a typical data structure like the data structure that presented in [16] is created that this node is its owner. If there are m sets of query points, this data structure has m lists. These lists are used for saving nodes of query points R tree used for finding nearest neighbor of the owner. So the search for nearest neighbor doesn't need to start from the root node of the query points R tree but can continue from saving nodes in lists of parent data structure. For each list in data structure of a node minimum of MinMindist and minimum of MaxMaxdist between MBR of owner node and MBRs of the nodes in list are stored.
Procedure of algorithm is that at first a new data structure is created that the root of data points R tree is its owner. In each m lists of this data structure, the root of each query points R tree is inserted. Same as BBS algorithm, a priority queue is used for storing and retrieving these data structures. After creating the root data structure, it is inserted to priority queue. The priority of each data structure determined with sum of minimums of MinMindists for all lists of it. The node that this value is lower for it is more promising. At each stage the data structure of the more promising node is removed from the priority queue and if it isn't dominated by skyline points have been found so far, it will be examined.

---

[1] minimum bounding rectangle

If the owner of this data structure is an intermediate node, a new data structure is created for each child of owner. For filling lists of child data structure, the nodes of corresponding lists of parent data structure are examined. If those nodes are leaves, themselves and if they are intermediate nodes, their children are inserted in the list of new data structure. Therefore by going down one level in data points R tree, we are going down one level in query points R trees too.

When a node is inserted in a list of data structure of child node the values of minimum of MinMindist and minimum of MaxMaxdist are updated for that list. When a list was filled with examination of all nodes in corresponding list of parent data structure, all nodes in the list are examined again and each node that isn't promising for finding nearest neighbor, is pruned. In the other words all nodes in list that minimum distance between their MBRs and owner MBR is greater than minimum of MaxMaxdist for that list, are removed from list. Because there is another node in the list that maximum distance between its MBR and owner MBR is less than minimum distance between this node and owner. So it is impossible that this node has nearest neighbor.

If the owner of new data structure is a leaf node, the exact distance of that node and its nearest neighbor from each set of query points should be calculated. So the procedure of filling lists with children of further nodes in lists and pruning none promising nodes should be continued until all nodes in lists are leaves and exact distances to nearest neighbors from each query sets are obtained.

After creating child data structure and filling its lists, it is examined that if this child is dominated by any skyline points found so far or not. For examining domination of a node, minimum of MinMindist for each list is used as an estimate of the distance to nearest neighbor. This is valid value for estimating distance to nearest neighbor because this is impossible that this value would be greater than distance to nearest neighbor. If child node isn't dominated, it is inserted to priority queue.

If the owner of the data structure that is removed from priority queue is a leaf node, this node is a skyline point and its data structure is inserted into list of skyline points. according to [4], because sum of minimums of MinMindists for all list is considered as promising criterion and this value is more optimistic than the sum of distances to nearest neighbors from each sets of query points, you can sure that this point is skyline and will not be dominated by any skyline points that found later. Search for finding skyline points is continued until the priority queue is empty.

By little changes a great improvement in algorithm can be gain. For high level nodes in data points R tree minimum of MinMindist for these node and the nodes of query points R tree is very small and in most cases because the intersection of these nodes, this value is zero. So this is not good criterion for promising nodes. If instead of this values, the sum of minimums MinMindists and minimums MaxMaxdists for all lists is used for promising criterion, better view of query point density around node is obtained. Only problem is that because this criterion is not more optimistic than exact distance to nearest neighbor, when a leaf node is removed from the priority queue, we can't tell this node is certainly a skyline point and the list of skyline points must be check and all nodes that are dominated with this node must be removed from the list. With this change in the algorithm, the number of nodes that inserted in and removed from priority queue is improved because better nodes are removed from priority queue and examined earlier.

The pseudo code of $N^2S^2$ algorithm can be seen below. This algorithm gives R tree of data points and R trees of each m sets of query points as parameters and returns a list of data structures of skyline points. Above-mentioned data structure is called DS in pseudo code. If we have m sets of query points, each DS has an owner and m lists. For each list the minimum of MinMindist and Minimum of MaxMaxdist is saved in the DS. These two values for list i'th called minminmindist(i) and minmaxmaxdist(i).

```
Algorithm N2S2(R-tree R, R-tree R_Q1 ... R-tree R_Qm)
1.  S=∅ //list of DSes that their owner is skyline point
2.  ds_root = New DS(R.root,m)
3.  for each R-tree R_Qi
4.      INSERT(ds_root, R_Qi.root, i)
5.  insert ds_root in the heap with priority PRI(ds_root,m)
6.  while heap not empty
7.      remove top DS ds
8.      if DOMINATE(S_i,ds,m) for some S_i in S discard ds
9.      else // ds is not dominated
10.         if ds.owner is an intermediate entry
11.             for each child ei of ds.owner
12.                 ds_ei = New DS(ei,m)
13.                 for all i from 1 to m
14.                     FILL(ds_ei,ds,i)
15.                 if DOMINATE(S_i,ds_ei,m) for some S_i in S discard ds_ei
16.                 else // ds_ei is not dominated
17.                     insert ds_ei in the heap with priority PRI(ds_ei,m)
18.         else // ds is a data point
19.             remove all S_i from S that DOMINATE(ds,DS_i,m)
20.             insert ds into S
21. end while
22. return S
23. EndN²S²
```

Figure 2. Pseudo code of N2S2 algorithm

This pseudo code has five functions. The INSERT function inserts a node into one of lists of a DS and updates minminmindist and minmaxmaxdist values corresponding to this list of DS. The PRI function calculates the priority of a DS. This function calculates the sum of all minminmindist and minmaxmaxdist values corresponding to all lists of the DS and returns this value. If the value of this function is less for a DS, it is more promising.

```
Algorithm INSERT(DS ds, Node n, Integer i)
//insert node n in i'th queue of data structure ds

 1.  ds.list(i).ADD(n)
 2.  temp= MinMin(ds.owner.MBR , n.MBR)
 3.  if temp < ds.minminmindist(i)
 4.      ds.minminmindist(i) = temp
 5.  temp= MaxMax(ds.owner.MBR , n.MBR)
 6.  if temp < ds.minmaxmaxdist(i)
 7.      ds.minmaxmaxdist(i) = temp
 8.  End INSERT
```

Figure 3.  INSERT function

```
Algorithm PRI(DS ds, Integer m)
//determine priority of the ds

 1.  Sum=0;
 2.  for all i from 1 to m
 3.      sum=sum + minminmindist(i) + minmaxmaxdist(i)
 4.  return sum
 5.  End PRI
```

Figure 4.  PRI function

The DOMINATE function gives two DS as input and examine if first DS dominate another or not. A DS dominate another DS if for all lists in this DS the value of minminmindist is less than or equal and for at least one list this value is less than the corresponding value of another list.

```
Algorithm DOMINATE(DS p, DS q, Integer m)
//if S_i dominate ds return true

 1.  check1=true
 2.  check2=false
 3.  for all i from 1 to m
 4.      if p.minminimndist(i)>q.minminmindist(i)
 5.          check1= false
 6.      else
 7.          if p.minminimndist(i)<q.minminmindist(i)
 8.              check2=true
 9.  return (check1 AND check2 )
10.  End DOMINATE
```

Figure 5.  DOMINATE function

```
Algorithm RECONFIG(DS ds, Integer i) //reconfigure queues of DS and prune

 1.  l=new list()
 2.  temp= ds.list(i).HEAD
 3.  while temp is not null
 4.      if MinMin(ds.owner.MBR , temp.MBR) < minmaxmaxdist(i)
 5.          l.ADD(temp)
 6.          temp=temp.NEXT
 7.  ds.list(i)=temp
 8.  End RECONFIG
```

Figure 6.  RECONFIG function

The RECONFIG function gives a DS and examines all nodes in one of its lists and removes all nodes from the list that are not promising for finding nearest neighbor of the owner. Each node that the minimum distance between it and the owner is greater than minmaxmaxdist of that list is not promising and removes from the list.

The FILL function fills a list of child DS according to corresponding list of parent DS. If the nodes in the list of parent DS are leaves, they are inserted in the list of child DS without any changes otherwise their children are inserted in the list of child DS. In both cases RECONFIG function is called for this list for pruning none promising nodes of DS. With the condition is in fourteen line of FILL function, a number of none promising nodes are pruned but because the value of minmaxmaxdist is not determined before complete filling of the list, a number of none promising nodes remain in the list and must be pruned with RECONFIG function. If the owner of child DS is leaf node exact distance of that node and its nearest neighbor should be calculated. So the FILL function is called recursively until all nodes in the list are leaves and exact distance to nearest neighbor is obtained.

```
Algorithm FILL(DS parent, DS child, Integer i)
//fill the i'th lists of one DS with another DS

1.  list = parent.list(i).COPY()
2.  child.list(i)=new LIST()
3.  child.minminmindist(i)=∞
4.  child.minmaxmaxdist(i)=∞
5.  if list contains leaf nodes
6.      temp= list.HEAD
7.      while temp is not null
8.          INSERT(child, temp, i)
9.          temp=temp.NEXT
10. else
11.     temp= list.HEAD
12.     while temp is not null
13.         for all child e of temp
14.             if MinMin(child.owner.MBR , temp.MBR)<child.minmaxmaxdist(i)
15.                 INSERT(child, e, i)
16.                 temp=temp.NEXT
17.             else
18.                 discard other childs of temp
19. RECONFIG(child,i)
20. If child.owner is leaf node
21.     While child.list(i) don't contains leaf nodes
22.         FILL(child,child,i)
23. End FILL
```

Figure 7. FILL functions

## 5. PERFORMANCE EVALUATION

Because spatial nearest neighbor skyline query for the first time is presented in this paper, for evaluation this algorithm, it is only compared to BBS algorithm for this kind of dynamic skyline query. There isn't any other algorithm to the best of our knowledge for this kind of query.

This algorithm was implemented with java programming language in NetBeans IDE. There were R* indexes in data and each set of query points. R* index was implemented with java and according to [17]. All indexes of data points and query points were stored in disk. For each index there was a catch memory for 512 numbers of tree blocks. The block length for all R* trees was

International Journal of Database Management Systems ( IJDMS ) Vol.3, No.4, November 2011

1K bytes. The experiments were performed on a computer with these specifications: pentume4, Intel, 3.06 GH, 512 MB of RAM and 80 GB of HDD.

This query has three parameters: the number of data points, the number of sets of query points and the number of query points in each set. The evaluations were performed in three phases and in each phase one of these parameters was changed and two other parameters were fixed. so each phase has some round of executing algorithms. For each round the new random sets of data points and query points were created and both algorithms executed for them.

Algorithms were compared in terms of CPU time and the number of IO. CPU time was calculated with one of library functions of java. This function calculates and returns the amount of time that the thread responsible for executing algorithms runs on CPU.

### 5.1. THE CHANGES IN THE NUMBER OF QUERY POINTS IN EACH SET

The first phase evaluated and compared algorithms related to changes in number of query points in each set. In this phase the number of data points and the number of sets of query points were fixed. There were 20000 data points and two sets of query points. The number of query points in each set was changed from 10000 to 25000. Figure 8 and 9 show results of comparing two algorithms. As can be seen in figures, with increasing the number of query points in each sets the chart of N2S2 algorithms are below BBS algorithm. So the CPU times and the number of IOs in N2S2 algorithm is less than BBS algorithm in all rounds.

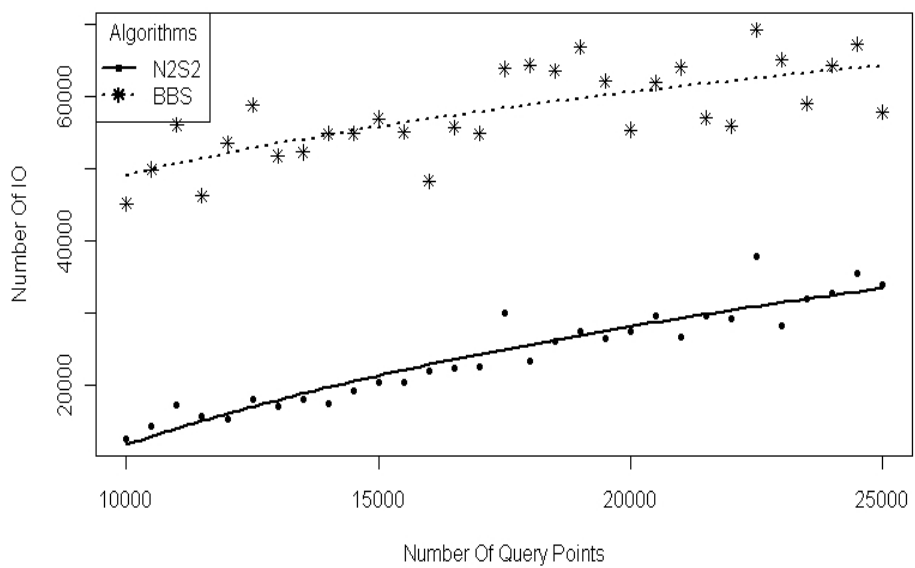

Figure 8. The number of IOs related to the number of query points in each set

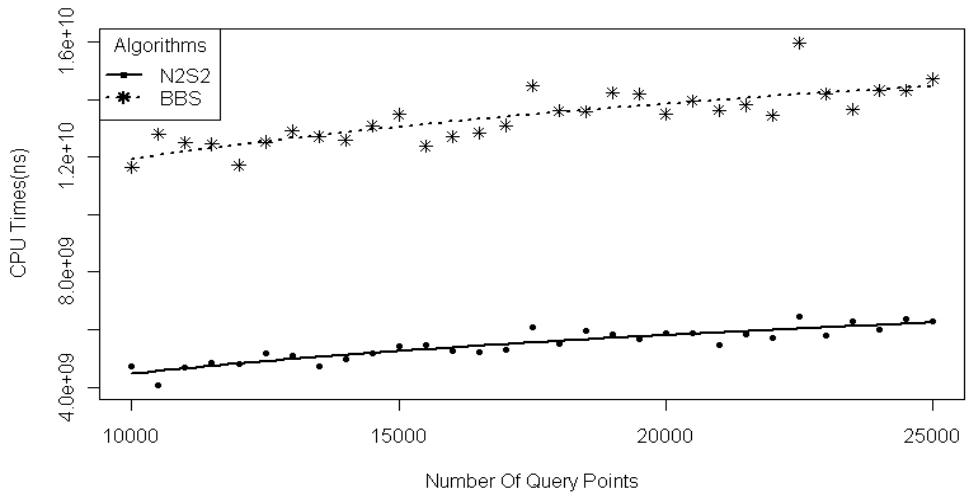

Figure 9. The CPU times related to the number of query points in each set

## 5.2. THE CHANGES IN THE NUMBER OF DATA POINTS

In the next evaluation the number of data points was changed and the number of sets of query points and points in each set were fixed. There were two sets of query points and 10000 query points in each set. The number of data points was changed from 10000 to 100000. Charts of this evaluation are seen in figures 10, 11 and 12.

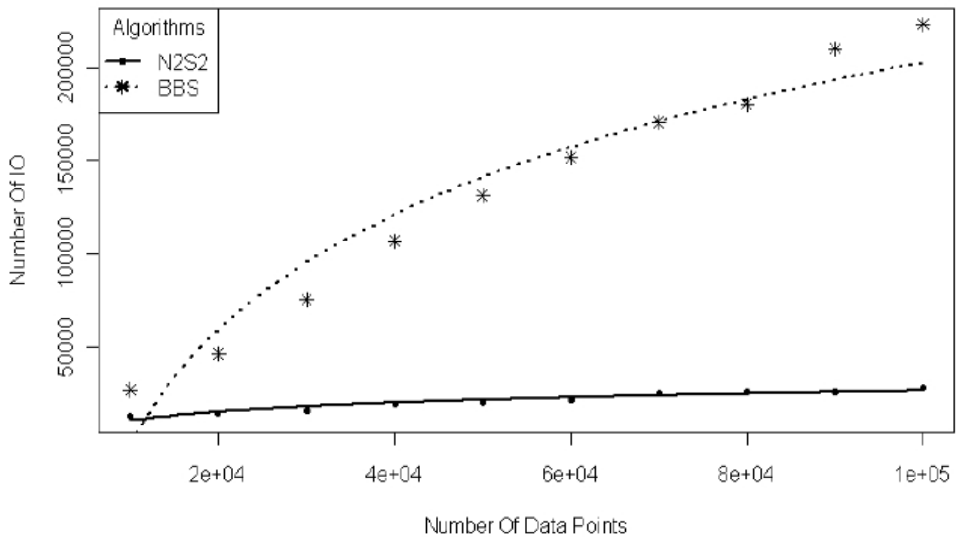

Figure 10. The number of IOs related to the number of data points

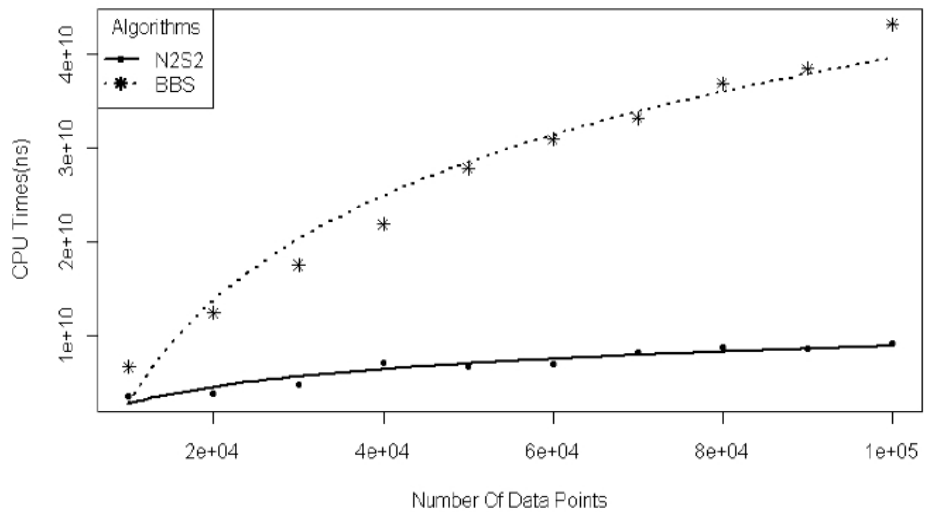

Figure 11. The CPU times related to the number of data points

As can be seen in figures the number of IOs and the CPU times in N2S2 algorithm is much less than BBS algorithm and the speed of growth in the charts of BBS algorithm is very fast.

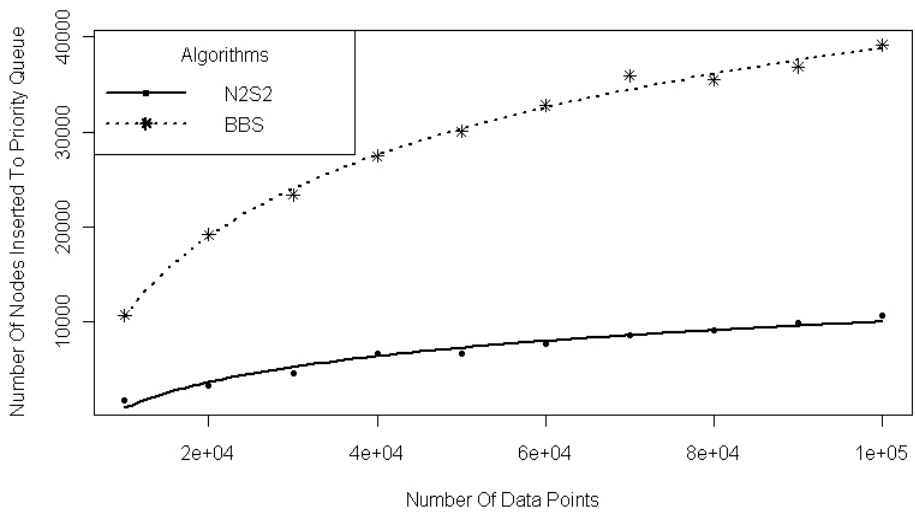

Figure 12. The number of nodes inserted in priority queue related to the number of data points

The number of nodes inserted in priority queue is important parameter because this parameter shows the performance of pruning metrics. Although the effect of pruning metrics reflects in the number of IOs and CPU time but probing this metric independently shows the efficiency of these metrics more obvious. This parameter is calculated and shows in figure 12. As this figure shows, the number of nodes inserted in priority queue in N2S2 algorithm is less than BBS algorithm.

**5.3. THE CHANGES IN THE NUMBER OF SETS OF QUERY POINTS**

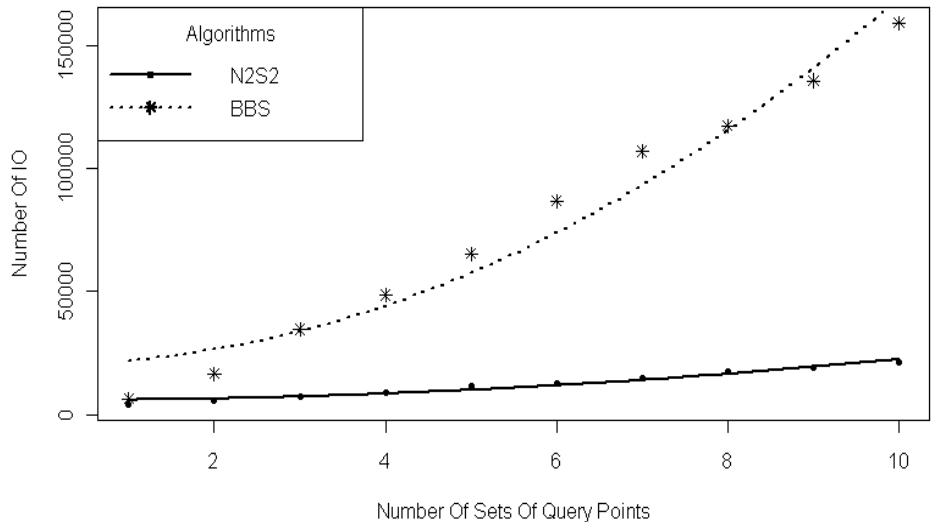

Figure 13. The number of IOs related to the number of sets of query points

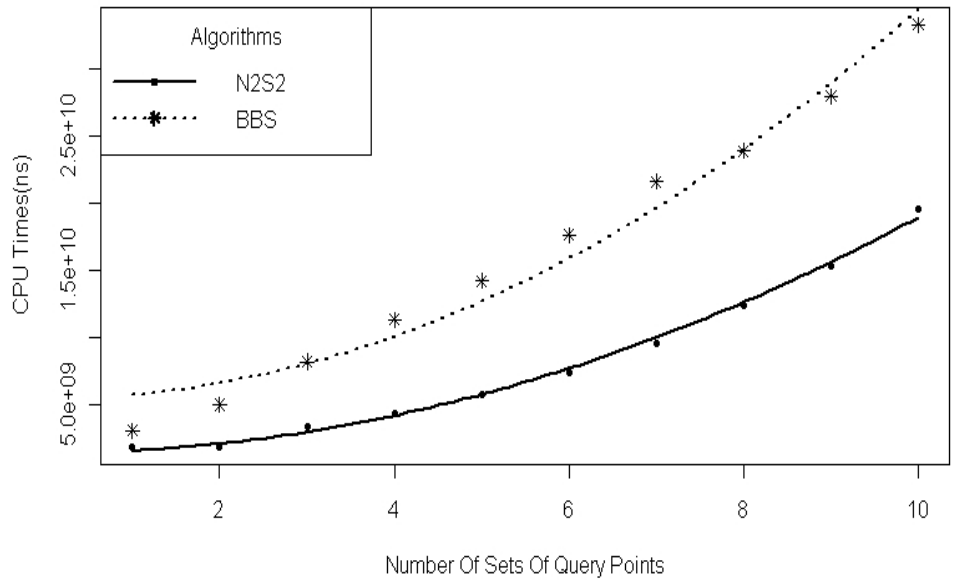

Figure 14. The CPU times related to the number of sets of query points

The last evaluation, probed and compared behaviour of algorithms related to changes in the number of sets of query points so two other parameters were fixed. figures 13 and 14 show that the number of IOs and CPU time in N2S2 algorithm is less than BBS algorithm.

## 6. CONCLUSIONS AND FUTURE WORKS

In this paper a new kind of user preference queries in spatial data bases is introduced that called spatial nearest neighbor skyline queries. This query is very practical in many fields such as service recommendation systems and investment planning. With separating it as a subset of dynamic skyline query, the $N^2S^2$ algorithm is presented for solving this kind of query with better performance. The number of IO in this algorithm is much less than BBS algorithm for this kind of queries. In future works we want to extend this kind of query and consider the quality of query points in addition to their distance from data point.

## Authors

**Nasrin Mazaheri Soudani** is M.Sc. student of software engineering at the Faculty of Engineering of the University Of Isfahan (UI). He earned his B.Sc. degrees from the University of Isfahan His research interest is spatial database systems.

**Ahmad Baraani-Dastjerdi** is an assistant professor of computer engineering at the School of Engineering of the University Of Isfahan (UI). He got his BS in Statistics and Computing in 1977. He got his M.Sc. & PhD degrees in Computer Science from George Washington University in 1979 &University of Wollongong in 1996, respectively. He is Head of the Research Department of the Communication systems and Information Security (CSIS) and Head of the ACM International Collegiate Programming Contest (ACM/ICPC) of University of Isfahan from 2000 until present. He co-authored three books in Persian and received an award of "the Best e-Commerce Iranian Journal Paper" (2005). Currently, he is teaching PhD and MS courses of Advance Topics in Database, Data Protection, Advance Databases, and Machining Learning. His research interests lie in Databases, Data security, Information Systems, e-Society, e-Learning, e-Commerce, Security in e-Commerce, and Security in e-Learning.